\documentclass{article}
\usepackage{bm}
\begin{document}

\title{Nonzero $\theta_{13}$, CP Violation, and  $\mu-\tau$ Symmetry}
\author{\bf{Asan Damanik}\footnote{E-mal: d.asan@lycos.com}\\Faculty of Science and Technology\\Sanata Dharma University\\Kampus III USD Paingan Maguwoharjo Sleman Yogyakarta\\Indonesia}
\date{}

\maketitle

\abstract{The nonzero and relatively large $\theta_{13}$  from the latest experimental results have a serious implication on the well-known neutrino mixing matrix.  One of the well-known mixing matrix is tribimaximal (TBM) neutrino mixing matrix which predict $\theta_{13}=0$. In order to accommodate nonzero $\theta_{13}$ and CP violation, we modified TBM by introducing a simple perturbation matrix into TBM matrix that can produces $\theta_{13}=7.89$ which is in agreement with the present experimental results.  The Dirac phase $\delta=77.20^{o}$ and the Jarlskog rephasing invariant: $J_{\rm CP}\approx 0.044$  are also obtained. The obtained neutrino mass matrix from the modified TBM with both nonzero $\theta_{13}$ and $\delta$ is the complex neutrino mass matrix.  If we impose the $\mu-\tau$ symmetry, as a constraint into neutrino mass matrix,one find that the Jarlskog rephasing invariant: $J_{\rm CP}=0$  which implies that CP violation cannot be accommodated in the $\mu-\tau$ symmetry scheme.
}

\section{Introduction}
Recent experimental results show that the mixing angle $\theta_{13}\neq 0$ and its value relatively large.  Nonzero $\theta_{13}$ have some implications in our understanding about the neutrino sector.  One of the implications when $\theta_{13}\neq 0$ is the possibility of CP violation in neutrino sector as well as quarks sector.   The nonzero and relatively large mixing angle $\theta_{13}$ have been reported by the follwing collaborations:  MINOS \cite{Minos}, Double Chooz \cite{Double}, T2K \cite{T2K},  Daya Bay  \cite{Daya}, and RENO \cite{RENO}.  From the theoretical side, three types of the well-known neutrino mixing matrices: tribimaximal (TBM)l, bimaxima (BM)l, and democratic (DM) and all of these three types of neutrino mixing matrices predict the  mixing angle $\theta_{13}=0$.  The evidence of nonzero $\theta_{13}$ due to the achievement of experimental methods and tools, the assumption that the value of mixing angle $\theta_{13}$ is zero must be corrected or even ruled out.  Concerning with the well-known mixing matrix, especially TBM, Ishimori and Ma \cite{Ishimori} stated explicitly that the tribimaximal mixing matrix may be dead due to the experimental fact that mixing angle $\theta_{13}\neq 0$. 

To explain the evidence of nonzero and relatively large $\theta_{13}$, several authors have already proposed some methods and models.   The simple way to accommodate a nonzero $\theta_{13}$ is to modify the neutrino mixing matrix by introducing a perturbation matrix into known mixing matrix such that it can produces a nonzero $\theta_{13}$ \cite{Boudjemaa, He11, Damanik, Damanik1, RodejohanW}, and the other is to build the model by using some discrete symmetries \cite{Cao, Luca}.  The nonzero $\theta_{13}$ is also known related to the Dirac phase $\delta$ as one can see in the standard parameterization of the neutrino mixing matrix.  Thus, nonzero $\theta_{13}$ gives a clue to the possible determination of CP violation in neutrino sector. Perturbation of neutrino mixing matrix in order to accommodate both nonzero $\theta_{13}$ and CP violation have been reported \cite{HarrisonScott, Harrison1, Grimus, Friedberg, Xing, Zhou}.

Concerning the $\mu-\tau$  symmetry and mixing angle $\theta_{13}$, Mohapatra \cite{Mohapatra} stated explicitly that neutrino mass matrix that obey $\mu-\tau$ symmetry to be the reason for maximal $\mu-\tau$ mixing, one gets $\theta_{13}=0$ and conversely if $\theta_{13}\neq 0$ can provide the $\mu-\tau$ symmetry beraking manifests in the case of normal hierarchy.  The nonzero $\theta_{13}$ and its implication to the leptogenesis as an origin of matter is discussed in \cite{Mohapatra1}.   Aizawa and Yasue \cite{Aizawa} analysis complex neutrino mass texture and the $\mu-\tau$ symmetry which can yield small $\theta_{13}$ as a $\mu-\tau$ breaking effect.   The $\mu-\tau$ symmetry breaking effect in relation with the small $\theta_{13}$ also discussed in \cite{Fuki}.  Another scenario that can produced Dirac CP phase by studying other mixing scenarios which deviate from tri-bimaximal mixing by leaving only one of the columns or one of the rows invariant which is called as "generalized trimaximal mixing" \cite{Albright, Albright1}.  Analysis of the correlation between CP violation and the $\mu-\tau$ symmetry breaking can be read in \cite{Mohapatra2, Baba}.  In \cite{Ge, Ge1}, the Dirac CP phace $\delta$ can be obtained by exploring the generalized $\bf {Z_{2}^{s}}$ symmetry without assumming $\mu-\tau$ symmetry.

In this paper we derive nonzero $\theta_{13}$ by modifying the TBM by introducing a simple perturbation matrix into TBM  and we then calculate the mixing angle $\theta_{13}$ by using the advantages of the mixing angles $\theta_{21}$ and $\theta_{32}$ from the experimental results.  The modified TBM to be used for obtaining the neutrino mass matrix $M_{\nu}$ in flavor basis and then we analysis the effect of $\mu-\tau$ symmetry on $M_{\nu}$.   This paper is organized as follow: in section 2, we modify TBM  by introducing a simple perturbation matrix.  In section 3, we determine the $\delta$ Dirac neutrino phase and neutrino masses constrained by $\mu-\tau$ symmetry.   Finally, section 4 is devoted to conclusion.

\section{Nonzero $\theta_{13}$ from the modified TBM}

The neutrino mixing matrix existence is due to the experimental facts that mixing of flavors do exist in the leptonic sector especially in  neutrino sector as well as in the quarks sector. The neutrino eigenstates in flavor basis ($\nu_{e}, \nu_{\mu}, \nu_{\tau}$) relate to the eigenstates of neutrino in mass basis ($\nu_{1}, \nu_{2}, \nu_{3}$) as follow:
\begin{eqnarray}
\nu_{i}=V_{ij}\nu_{j}
\end{eqnarray}
where $V_{ij} ( i = e, \mu, \tau; j = 1,2,3)$ are the elements of neutrino mixing matrix.  The standard parameterization of the mixing matrix $V$  read:
\begin{eqnarray}
V=\bordermatrix{& & &\cr
&c_{12}c_{13} &s_{12}c_{13} &s_{13}e^{-i\delta}\cr
&-s_{12}c_{23}-c_{12}s_{23}e^{i\delta} &c_{12}c_{23}-s_{12} s_{23}e^{i\delta}&s_{23}c_{13}\cr
&s_{12}s_{23}-c_{12}c_{23}e^{i\delta} &-c_{12}s_{23}-s_{12}c_{23}e^{i\delta} &c_{23}c_{13}}
 \label{V}
\end{eqnarray}
where $c_{ij}$ is the $\cos\theta_{ij}$, $s_{ij}$ is the $\sin\theta_{ij}$, and $\theta_{ij}$ are the mixing angles.

One of the well-known neutrino mixing matrix ($V$) is the TBM ($V_{TBM}$) which given by \cite{Harrison, Harrisona, Xing1, Harrisonb, Harrisonc, He}:
\begin{equation}
V_{TBM}=\bordermatrix{& & &\cr
&\sqrt{\frac{2}{3}} &\frac{1}{\sqrt{3}} &0\cr
&-\frac{1}{\sqrt{6}} &\frac{1}{\sqrt{3}} &\frac{1}{\sqrt{2}}\cr
&-\frac{1}{\sqrt{6}} &\frac{1}{\sqrt{3}} &-\frac{1}{\sqrt{2}}}.
 \label{tb}
\end{equation}
As one can see from Eq. (\ref{tb}) that the entry $V_{e3}=0$ which imply that the mixing angle $\theta_{13}$ must be zero in the TBM.   However, the latest result from long baseline neutrino oscillation experiment T2K indicates that $\theta_{13}$ is nonzero and  relatively large.  For a vanishing Dirac CP-violating phase ($\delta = 0$), the T2K collaboration reported that the values of $\theta_{13}$ for neutrino mass in normal hierarchy (NH) are \cite{T2K}:
\begin{equation}
5.0^{o}\leq\theta_{13}\leq 16.0^{o},
\end{equation}
and for neutrino mass in inverted hierarchy (IH):
\begin{equation}
5.8^{o}\leq\theta_{13}\leq 17.8^{o},
\end{equation}
Tthe current combined world data for neutrino squared-mass difference are given by \cite{Gonzales-Carcia, Fogli}:
\begin{equation}
\Delta m_{21}^{2}=7.59\pm0.20 (_{-0.69}^{+0.61}) \times 10^{-5}~\rm{eV^{2}},\label{21}
\end{equation}
\begin{equation}
\Delta m_{32}^{2}=2.46\pm0.12(\pm0.37) \times 10^{-3}~\rm{eV^{2}},~\rm(for~ NH)\label{32}
\end{equation}
\begin{equation}
\Delta m_{32}^{2}=-2.36\pm0.11(\pm0.37) \times 10^{-3}~\rm{eV^{2}},~\rm(for~ IH)\label{321}
\end{equation}
\begin{equation}
\theta_{12}=34.5\pm1.0 (_{-2.8}^{3.2})^{o},~~\theta_{23}=42.8_{-2.9}^{+4.5}(_{-7.3}^{+10.7})^{o},~~\theta_{13}=5.1_{-3.3}^{+3.0}(\leq 12.0)^{o},
 \label{GD}
\end{equation}
at $1\sigma~(3\sigma)$ level.  The latest experimental result for the value of $\theta_{13}$ is reported by Daya Bay Collaboration which gives \cite{Daya}:
\begin{equation}
\sin^{2}2\theta_{13}=0.092\pm 0.016 (\rm{stat}.)\pm 0.005 (\rm{syst.}),
\end{equation}
and RENO Collaboration reported that \cite{RENO}:
\begin{equation}
\sin^{2}2\theta_{13}=0.113\pm 0.013 (\rm{stat.})\pm 0.014 (\rm{syst.}).
\end{equation}

Concerning the TBM status in the context of nonzero $\theta_{13}$, Ishimori and Ma \cite{Ishimori} concluded that TBM may be dead but $A_{4}$ alive and even getting healthier.   Modification of neutrino mixing matrix, by introducing  a perturbation matrix into neutrino mixing matrix in Eq. (\ref{tb}), is the simple way to obtain the nonzero $\theta_{13}$.  The value of $\theta_{13}$ can be obtained in some parameters that can be fitted from experimental results.  Nonzero of $V_{e3}$ for TBM can be obtained by charge lepton corrections and renormalization group running \cite{Boudjemaa}.  In this paper, the modified TBM neutrino mixing matrix to be considered are given by:
\begin{equation}
V_{{\rm TBM}}^{'}=V_{{\rm TBM}}V_{y},\label{Modi1}
\end{equation}
where $V_{y}$ is the perturbation matrix.  We take the form of the perturbation matrix as follow:
\begin{equation}
V_{y}=\bordermatrix{& & &\cr
&1 &0 &0\cr
&0 &c_{y} &s_{y}e^{-i\delta}\cr
&0 &-s_{y}e^{i\delta} &c_{y}\cr}.
 \label{xy}
\end{equation}
where $c_{y}$ is the $\cos{y}$, $s_{y}$ is the $\sin{y}$, and $\delta$ is the Dirac CP phase.

By inserting Eqs. (\ref{tb}) and (\ref{xy}) into Eqs. (\ref{Modi1}), we then have the modified neutrino mixing matrix as follow:
\begin{equation}
V_{{\rm TBM}}^{'}=\bordermatrix{& & &\cr
&\frac{\sqrt{6}}{3} &\frac{\sqrt{3}}{3}c_{y} &\frac{\sqrt{3}}{3}s_{y}e^{-i\delta}\cr
&-\frac{\sqrt{6}}{6} &\frac{\sqrt{3}}{3}c_{y}-\frac{\sqrt{2}}{2}s_{y}e^{i\delta} &\frac{\sqrt{3}}{3}s_{y}e^{-i\delta}+\frac{\sqrt{2}}{2}c_{y}\cr
&-\frac{\sqrt{6}}{6} &\frac{\sqrt{3}}{3}c_{y}+\frac{\sqrt{2}}{2}s_{y}e^{i\delta} &\frac{\sqrt{3}}{3}s_{y}e^{-i\delta}-\frac{\sqrt{2}}{2}c_{y}},\label{Mo1}
\end{equation}
By comparing Eqs. (\ref{Mo1}) with the neutrino mixing in standard parameterization form as shown in Eq. (\ref{V}), we have:
\begin{equation}
\tan\theta_{12}=\left|\frac{\sqrt{2}c_{y}}{2}\right|,~~
\tan\theta_{23}=\left|\frac{\frac{\sqrt{3}}{3}s_{y}e^{-i\delta}+\frac{\sqrt{2}}{2}c_{y}}{\frac{\sqrt{3}}{3}s_{y}e^{-i\delta}-\frac{\sqrt{2}}{2}c_{y}}\right|,~~
\sin\theta_{13}=\left|\frac{\sqrt{3}}{3}s_{y}e^{-i\delta}\right|,
 \label{1}
\end{equation}
and for $\delta=0$ \cite{Damanik1}:
\begin{equation}
\tan\theta_{12}=\left|\frac{\sqrt{2}c_{y}}{2}\right|,~~
\tan\theta_{23}=\left|\frac{\frac{\sqrt{3}}{3}s_{y}+\frac{\sqrt{2}}{2}c_{y}}{\frac{\sqrt{3}}{3}s_{y}-\frac{\sqrt{2}}{2}c_{y}}\right|,~~
\sin\theta_{13}=\left|\frac{\sqrt{3}}{3}s_{y}\right|.
 \label{2}
\end{equation}

From Eqs. (\ref{1}) and (\ref{2}) it is apparent that for $y\rightarrow 0$, the value of $\tan\theta_{12}\rightarrow \sqrt{2}/2$  and $\tan\theta_{23}\rightarrow 1$ which imply that $\theta_{12}\rightarrow 35.264^{o}$ and $\theta_{23}\rightarrow 45^{o}$.   From Eq. (\ref{1}), one can see that it is possible to determine the value $y$ and therefore the value of $\theta_{13}$ by using the experimental values of $\theta_{12}$ and $\theta_{23}$ in Eq. (\ref{GD}).

By inserting the experimental values of $\theta_{12}$ and $\theta_{23}$ in Eq. (\ref{GD}) into Eq. (\ref{2}), we obtain the relations:
\begin{equation}
c_{y}=-0.03167630078 s_{y},\label{c1}
\end{equation}
when we use $\theta_{23}$, and
\begin{equation}
c_{y}=0.9713265692,\label{c2}
\end{equation}
when we use $\theta_{12}$.  From both Eqs. (\ref{c1}) and (\ref{c2}), we can see that the realistic value for $c_{y}$ is the value of the $c_{y}$ in Eq. (\ref{c2}) which gives $y=13.7537^{o}$.  It means that in this modification scenario, only the experimental mixing angle $\theta_{12}$ related to the mixing angle $\theta_{13}$.  From Eq. (\ref{c2}), we have:
\begin{eqnarray}
\sin\theta_{13}=\sqrt{1-c_{y}^{2}}=0.137265,
\end{eqnarray}
that imply the mixing angle $\theta_{13}=7.89^{o}$ which is in agreement with the T2K \cite{T2K} and Daya Bay experimental results \cite{Daya}.

It is also apparent from Eq. (\ref{1} )that only the mixing angle $\theta_{23}$ related to the Dirac phase $\delta$.  Thus, it is possible to detect the Dirac phase $\delta$ by measuring the precise value of mixing angle $\theta_{23}$ in the future experiment.  If we substitute the value of $\theta_{23}=42.8^{o}$ as shown in Eq. (\ref{GD}) into Eq. (\ref{1}), then we have:
\begin{equation}
\delta=77.20^{o}.\label{delta}
\end{equation}

The Jarlskog invariant $J_{\rm CP}$, which is very uselful for quantifying CP violation, is given by \cite{Branco}:
\begin{equation} 
J_{\rm CP}={\rm Im}\left[(V_{\rm TBM}^{'})_{11}(V_{\rm TBM}^{'})_{22}(V_{\rm TBM}^{'})_{12}^{*}(V_{\rm TBM}^{'})_{21}^{*}\right].\label{CP}
\end{equation}
If we inser the corresponding values of $V_{\rm TBM}^{'}$ of Eq. (\ref{Mo1}) into Eq. (\ref{CP}) with the values of $c_{y}$ in Eq. (\ref{c2}) and $\delta$ in Eq. (\ref{delta}), then we have:
\begin{equation}
J_{\rm CP}=0.043853.
\end{equation}

\section{$\mu-\tau$ symmetry and Dirac phase $\delta$}

Concerning the $\mu-\tau$ symmetry, most of the authors discuss the $\mu-\tau$ symmetry in relation with the mixing angle $\theta_{13}$ \cite{Mohapatra} and its implication to origin of matter via leptogenesis \cite{Mohapatra1}. The effect of $\mu-\tau$ symmetry broken in the neutrino mass matrix can arises the leptonic CP violation was proposed by Mohaptara and Rodejohann \cite{Mohapatra2}.  In this section, we analysis the $\mu-\tau$ symmetry in relation with the Dirac phase $\delta$.  To begin with the assumption that the charged lepton mass matrix is diagonal, then we evaluate the neutrino mass matrix $M_{\nu}$ in flavor eigenstate basis.  In the basis where the charged lepton mass matrix is diagonal, the neutrino mass matrix can be diagonalized by a unitary matrix $V$ as follow:
\begin{equation}
M_{\nu}=VMV^{T},
\label{MM}
\end{equation}
where the diagonal neutrino mass matrix $M$ is given by:
\begin{equation}
M=\bordermatrix{& & &\cr
&m_{1} &0 &0\cr
&0 &m_{2} &0\cr
&0 &0 &m_{3}}.
\label{NN}
\end{equation}

If the unitary matrix $V$ is the modified TBM  mixing matrix ($V'_{TBM}$) as shown in Eq. (\ref{Mo1}), then from Eq. (\ref{MM}) the neutrino mass matrix in favor basis is given by:
\begin{equation}
M_{\nu}=\bordermatrix{& & &\cr
&A &B &C\cr
&B &D &E\cr
&C &E &F},\label{Mv}
\end{equation}
where:
\begin{equation}
A=\frac{2m_{1}}{3}+\frac{m_{2}}{3}c_{y}^{2}+\frac{m_{3}}{3}s_{y}^{2}e^{-2i\delta},\label{M11}
\end{equation}
\begin{equation}
B=-\frac{m_{1}}{3}+m_{2}\left(\frac{1}{3}c_{y}^{2}-\frac{\sqrt{6}}{6}c_{y}s_{y}e^{i\delta}\right)+m_{3}\left(\frac{1}{3}s_{y}^{2}e^{-2i\delta}+\frac{\sqrt{6}}{6}s_{y}c_{y}e^{-i\delta}\right),\label{M12}
\end{equation}
\begin{equation}
C=-\frac{m_{1}}{3}+m_{2}\left(\frac{1}{3}c_{y}^{2}+\frac{\sqrt{6}}{6}c_{y}s_{y}e^{i\delta}\right)+m_{3}\left(\frac{1}{3}s_{y}^{2}e^{-2i\delta}-\frac{\sqrt{6}}{6}s_{y}c_{y}e^{-i\delta}\right),\label{M13}
\end{equation}
\begin{equation}
D=\frac{m_{1}}{6}+m_{2}\left(\frac{\sqrt{3}}{3}c_{y}-\frac{\sqrt{2}}{2}s_{y}e^{i\delta}\right)^{2}+m_{3}\left(\frac{\sqrt{3}}{3}s_{y}e^{-i\delta}+\frac{\sqrt{2}}{2}c_{y}\right)^{2},\label{M14}
\end{equation}
\begin{equation}
E=\frac{m_{1}}{6}+m_{2}\left(\frac{1}{3}c_{y}^{2}-\frac{1}{2}s_{y}^{2}e^{2i\delta}\right)+m_{3}\left(\frac{1}{3}s_{y}^{2}e^{-2i\delta}-\frac{1}{2}c_{y}^{2}\right),\label{M15}
\end{equation}
\begin{equation}
F=\frac{m_{1}}{6}+m_{2}\left(\frac{\sqrt{3}}{3}c_{y}+\frac{\sqrt{2}}{2}s_{y}e^{i\delta}\right)^{2}+m_{3}\left(\frac{\sqrt{3}}{3}s_{y}e^{-i\delta}-\frac{\sqrt{2}}{2}c_{y}\right)^{2},\label{M16}
\end{equation}

As one know that the TBM neutrino mixing matrix lead to the neutrino mass matrix with $\mu-\tau$ symmetry as the underlying symmetry, then in the rest of the remaining this paper we also impose the $\mu-\tau$ symmetry as a constraint to the obtained neutrino mass matrix $M_{\nu}$ from the modified TBM.   By imposing $\mu-\tau$ symmetry into neutrino mass matrix in Eq. (\ref{Mv}) it then imply : $B=C$ and $D=F$.

For $B=C$ and $D=F$ constraints, we have the relation:
\begin{equation}
\frac{m_{2}}{m_{3}}=e^{-2i\delta}.\label{m1m2}
\end{equation}
By inserting Eq. (\ref{m1m2}) into Eq. (\ref{Mv}), we then have the neutrino mass matrix in flavor basis as follow:
\begin{equation}
M_{\nu}=\bordermatrix{& & &\cr
&P &Q &Q\cr
&Q &R &S\cr
&Q &S &R},\label{Mv1}
\end{equation}
where:
\begin{equation}
P=\frac{1}{3}\left(2m_{1}+m_{2}\right),\label{P}
\end{equation}
\begin{equation}
Q=\frac{1}{3}(m_{2}-m_{1}),\label{Q}
\end{equation}
\begin{equation}
R=\frac{1}{6}\left(m_{1}+m_{2}(2+3e^{2i\delta})\right),\label{R}
\end{equation}
\begin{equation}
S=\frac{1}{6}\left(m_{1}+m_{2}(2-3e^{2i\delta})\right).\label{S}
\label{P}
\end{equation}

The effect of $\mu-\tau$  symmetry on neutrino mass matrix of Eq. (\ref{Mv}) is to reduce the number of parameters including the parameters that introduced in perturbation mixing matrix.  From three parameters ($c_{y}, s_{y}, \delta$) in the perturbation mixing matrix of Eq. (\ref{xy}), only the Dirac phase parameter $\delta$ appears when the neutrino mass matrix in Eq. (\ref{Mv1}) constrained by $\mu-\tau$ symmetry.  The eigenvalues of the neutrino mass matrix of Eq. (\ref{Mv1}) are:
\begin{equation}
\lambda_{1}=R-S,\label{ev1}
\end{equation}
\begin{equation}
\lambda_{2}=\frac{1}{2}\left(P+R+S-\sqrt{P^{2}-2RP-2PS+R^{2}+2RS+S^{2}+8Q^{2}}\right),\label{ev2}
\end{equation}
\begin{equation}
\lambda_{3}=\frac{1}{2}\left(P+R+S+\sqrt{P^{2}-2RP-2PS+R^{2}+2RS+S^{2}+8Q^{2}}\right).
\label{ev3}
\end{equation}
From Eqs. (\ref{ev1}), (\ref{ev2}), and (\ref{ev3}), we have:
\begin{equation}
\lambda_{1}+\lambda_{2}+\lambda_{3}=2R+P.
\end{equation}

The Jarlskog rephasing invariant $J_{\rm CP}$  can  be also determined from relation \cite{Branco}:
\begin{equation}
J_{\rm CP}=-\frac{{\rm Im}\left[(M_{\nu}^{'})_{e\mu}(M_{\nu}^{'})_{\mu\tau}(M_{\nu}^{'})_{\tau e}\right]}{\Delta m_{21}^{2}\Delta m_{32}^{2}\Delta m_{31}^{2}}=-\frac{{\rm Im}\left[(QQ^{\dagger})^2SS^{\dagger}\right]}{\Delta m_{21}^{2}\Delta m_{32}^{2}\Delta m_{31}^{2}}.\label{J}
\end{equation}
where $(M_{\nu}^{'})_{ij}=(M_{\nu}M_{\nu}^{\dagger})_{ij}$ and $ i,j=e,\nu,\tau$.  For the neutrino mass matrix constrained by $\mu-\tau$ symmetry, the value of $J_{\rm CP}$ can be obtained by inserting the values of $Q$ and $S$ in Eqs. (\ref{Q}) and (\ref{S}) into Eq. (\ref{J}) which gives:
\begin{equation}
J_{\rm CP}=0.\label{J1}
\end{equation}
From Eq. (\ref{J1}), one can see that the exact $\mu-\tau$ symmetry lead to $J_{\rm CP}=0$ as claimed by many authors.  If one still want to get the value of $J_{\rm CP}\neq 0$ in the scheme of $\mu-\tau$ symmetry, one must break the $\mu-\tau$ symmetry softly by introducing some parameters to break the neutrino mass matrix.

\section{Conclusion}
The nonzero and relatively large $\theta_{13}$  from the latest experimental results have a serious implication on the well-known neutrino mixing matrix.  One of the well-known mixing matrix is tribimaximal (TBM) neutrino mixing matrix which predict $\theta_{13}=0$. We modified TBM by introducing a simple perturbation matrix into TBM matrix and using the advantage of the experimental results of the mixing angles, we then can obtain $\theta_{13}=7.89$ which is in agreement with the present experimental results.  The Dirac phase $\delta=77.20^{o}$ and the Jarlskog rephasing invariant: $J_{\rm CP}\approx 0.044$  are also obtained. The obtained neutrino mass matrix from the modified TBM with both nonzero $\theta_{13}$ and $\delta$ is the complex neutrino mass matrix.  If we impose the $\mu-\tau$ symmetry, as a constraint into neutrino mass matrix,one find that the Jarlskog rephasing invariant: $J_{\rm CP}=0$  which implies that CP violation cannot be accommodated in the $\mu-\tau$ symmetry scheme.  If one still want to usee the $\mu-\tau$ symmetry for as the undelying symmetry for neutrino mass matrix, the one must break the $\mu-\tau$ symmetry softly.

\end{document}